\pdfoutput=1
\documentclass[aps,pra,reprint,groupedaddress]{revtex4-2} 
\usepackage{amsmath}
\usepackage{amsfonts}
\usepackage{graphicx}
\usepackage{natbib}
\usepackage{bm}
\usepackage[
colorlinks=true, linkcolor=blue, urlcolor=blue, citecolor=blue, bookmarks=false
]{hyperref}

\begin{document}

\title{Testing precision and accuracy of weak value measurements in an IBM quantum system}
\author{David R. A. Ruelas Paredes}
\email{druelas@pucp.edu.pe}
\author{Mariano Uria}
\altaffiliation[Present address: ]{Departamento de F\'isica, Facultad de Ciencias F\'isicas y Matem\'aticas, Universidad de Concepci\'on, Concepci\'on, Chile.}
\author{Eduardo Massoni}
\author{Francisco De Zela}
\email{fdezela@pucp.edu.pe}
\affiliation{Departamento de Ciencias, Secci\'on F\'isica \\ Pontificia Universidad Cat\'olica del Per\'u\\ Lima 15088, Peru}

\begin{abstract}
Historically, weak values have been associated with weak measurements performed on quantum systems. Over the past two decades, a series of works have shown that weak values can be determined via measurements of arbitrary strength. One such proposal by Denkmayr \textit{et~al.} [\href{https://doi.org/10.1103/PhysRevLett.118.010402}{Phys. Rev. Lett. \textbf{118}, 010402 (2017)}], carried out in neutron interferometry experiments, yielded better outcomes for strong than for weak measurements. {We extend this scheme and explain how to implement it in an optical setting as well as in a quantum computational context.} Our implementation in a quantum computing system provided by IBM confirms that weak values can be measured, with varying degrees of performance, over a range of measurement strengths. However, at least for this model, strong measurements do not always perform better than weak ones.
\end{abstract}

\maketitle

\section{\label{sec:introduction}Introduction}

Quantum weak values are complex numbers associated to an operator, an initial (preselection) state, and a final (postselection) state. More concretely, let $S$ be the quantum system of interest, $A_S$ an observable, $|i_S\rangle$ the initial state, and $|f_S\rangle$ the final state. The weak value of $A_S$ for these pre- and postselection states is defined as
\begin{equation}
	\label{eq:weakValueDefinition}
	\left\langle A_S\right\rangle_w \equiv \frac{\langle f_S|A_S|i_S\rangle}{\langle f_S|i_S\rangle}.
\end{equation}
Evidently, there is nothing inherently ``weak'' about the above quantity. A more descriptive label, such as $\langle A_S\rangle_{fi}$, would have the benefit of suggesting the natural interpretation as a generalized mean value. The adjective ``weak'' dates back to the introduction of weak values in the context of weak measurements \cite{aharonov1988}. To make sense of this qualifier, let us consider von Neumann's model of quantum measurements, wherein $P$ is the pointer (an ancillary system that represents the measurement apparatus), and $H_\text{int}=gA_S\otimes B_P=gA_SB_P$ describes the coupling between $S$ and $P$ while $A_S$ is being measured. In this interaction Hamiltonian, $B_P$ is a momentum operator, so it is conjugate to the position of the pointer space: $\left[X_P,B_P\right]=i$, with $\hbar=1$ and $X_P|x\rangle=x|x\rangle$; $g$ accounts for the coupling between both systems, and may be considered a function of time. A measurement of $A_S$ is characterized by a shift in the pointer's position. Indeed, let $|i_S\rangle= \sum_n\alpha_n|a_n\rangle$ be the state of the system, with $A_S|a_n\rangle=a_n|a_n\rangle$, and $|i_P\rangle=|\Phi(0)\rangle$ be the state of the pointer, a Gaussian state centered at $x=0$ with spread $\sigma$:
\begin{equation*}
\langle x|\Phi(0)\rangle= \phi(x) =(2\pi\sigma^2)^{-1/4} \exp\left(-x^2/2\sigma^2\right).
\end{equation*}
When $S$ and $P$ interact, their states evolve jointly under the action of
\begin{equation}\label{eq:evolutionOperator}
U(\epsilon)\equiv\exp\!\left( -i\int_0^{\Delta t} \mathrm{d}t'\,H_\text{int} \right) = \exp\left(-i\epsilon A_SB_P\right).
\end{equation}
The parameter $\epsilon\equiv \int_0^{\Delta t}g\,\mathrm{d}t'$, which we will assume positive, quantifies the interaction strength, whence the ``weak'' or ``strong'' character of a measurement can be ascertained. It can be shown \cite{[][{ (to be published).}]ruelas2023} that the system-pointer state then becomes 
\begin{equation}
U(\epsilon)|i_S\rangle|i_P\rangle= \sum_n \alpha_n |a_n\rangle |\Phi({\epsilon a_n})\rangle,
\end{equation}
a superposition of eigenstates of $A_S$ coupled to Gaussian states centered at $x=\epsilon a_n$, each with spread $\sigma$. {If $\epsilon(a_{n+1}-a_n)\gg\sigma$,} the pointer wave packets do not overlap, so a reading of the pointer's position $X_P$ reveals the measured eigenvalue of $A_S$ with no ambiguity; such a measurement is called strong. {For $\epsilon(a_{n+1}-a_n) \not\gg \sigma$,} conversely, the Gaussians in the evolved state do overlap, and the pointer's position is not unambiguously correlated with any single $a_n$; these are the so-called weak measurements.

Immediately after a weak measurement, the evolved state can be projected onto a final system state. This procedure, called postselection, yields the (unnormalized) final pointer state $|f_P\rangle= \langle f_S| U(\epsilon) |i_S\rangle |i_P\rangle$. Since the evolution operator of a weak measurement can be approximated as $U(\epsilon)=1_{SP}-i\epsilon A_SB_P+\mathcal{O}\left(\epsilon^2\right)$, the final pointer state reads
\begin{equation}
    \label{eq:finalPointerState}
    |f_P\rangle=\langle f_S|i_S\rangle\left(1_P-i\epsilon\left\langle A_S\right\rangle_wB_P+\mathcal{O}\!\left(\epsilon^2\right)\right)\!\int\!\mathrm{d}p\,\Tilde{\phi}(p)\,|p\rangle,
\end{equation}
where $1_{SP}$ and $1_P$ are the identity operators of the system-pointer and pointer spaces, respectively, and we have introduced the momentum representation of $|\Phi(0)\rangle$. Under suitable weakness conditions \cite{duck1989, ruelas2023}, the truncated sum in Eq.~(\ref{eq:finalPointerState}) can be rewritten as an exponential. In this approximation, after an inverse Fourier transform, $|f_P\rangle$ takes the form of a Gaussian centered at the complex number $\epsilon \!\left\langle A_S\right\rangle_w$:
\begin{align}
    \label{eq:ComplexCenteredGaussian}
    |f_P\rangle &\approx\langle f_S|i_S\rangle \int\!\mathrm{d}p\,\exp\!\left(-i\epsilon \langle A_S\rangle_w\,p\right)\Tilde{\phi}(p) \,|p\rangle\nonumber\\
    &=\langle f_S|i_S\rangle \int\!\mathrm{d}x\,\phi\!\left(x-\epsilon \langle A_S\rangle_w \right) |x\rangle\nonumber\\
    &=\langle f_S|i_S\rangle|\Phi\!\left(\epsilon \langle A_S\rangle_w\right)\rangle.
\end{align}
However, as it turned out, weak measurements are not a necessary condition for so-called weak values to emerge. Without need for the weakness hypothesis, Johansen \cite{johansen2007} proposed a way to measure the real and imaginary parts of a weak value; Zou \textit{et~al.} \cite{zou2015} presented an algorithm  that uses experimentally obtained weak values to determine the complex amplitudes of pure states; and Zhang \textit{et~al.} developed a theoretical framework wherein, by probing pointer observables that are functions of the interaction strength, weak values \cite{zhang2016pointerObservables} {as well as mixed states \cite{zhu2016direct} can be reconstructed}. Along these lines, Denkmayr \textit{et~al.} \cite{denkmayr2017, denkmayr2018} put forward a model for measuring the weak values of a system qubit's Pauli spin operator $\sigma_z$ via measurements of arbitrary strength on a pointer qubit. In neutron interferometry experiments, they measured $\left\langle\sigma_z\right\rangle_w$ once in a weak setting and once in a strong setting, and concluded that ``experimental evidence is given that strong interactions are superior to weak interactions in terms of accuracy and precision, as well as required measurement time'' \cite{denkmayr2018}. {Later experimental works \cite{calderaro2018, xu2021} arrived at similar conclusions, which had been first surmised from Vallone and Dequal's numerical simulations \cite{vallone2016}. In contrast, a series of theoretical results prior to Denkmayr \textit{et~al.}'s papers had shown that precision and accuracy are not always maximized at higher coupling strengths \cite{das2014estimation, gross2015, zhu2016direct}.}

For the present work, we extended Denkmayr \textit{et~al.}'s scheme in order to measure weak values of Pauli spin operators ${\sigma}_\mathbf{n}$, with $\mathbf{n}$ a real unit vector. Prompted by their conclusions, we set out to test our proposal across a wider range of system-pointer interaction strengths, for which we recurred to cloud-based quantum computation systems provided by the IBM Corporation. {Since these systems can be operated remotely via Python scripts and allow for the experimental setup to be varied effortlessly, they constitute a more versatile experimental context than neutron interferometry and optical settings.} Their basic building blocks are so-called superconducting transmon qubits, which are acted upon by quantum gates (operators). Made of materials such as niobium and aluminum, these qubits are kept in a dilution refrigerator at 15 mK with the goal of minimizing environmental effects \footnote{IBM Quantum, The qubit, currently available at \url{https://web.archive.org/web/20230605160728/https://quantum-computing.ibm.com/composer/docs/iqx/guide/the-qubit}}. In the Bloch-Redfield model, times $T_1$ and $T_2$ characterize the time scale in which a qubit, on average, undergoes decoherence \cite{krantz2019}. To successfully realize quantum circuits (sequences of operators), a qubit state must be preserved for sufficient time for the required gates to be implemented and measurement results to be read. From 2012 on, IBM's quantum systems achieved decoherence times that enabled useful computations \cite{bozzoRey2018}. Nowadays, the IBM Quantum platform gives free access to quantum systems. Since its release in 2016, researchers have employed the platform as a test tool for multiple experimental proposals, such as tests of Mermin inequalities \cite{alsina2016}; protocols for ``quantum error correction, quantum arithmetic, quantum graph theory, and fault-tolerant quantum computation'' \cite{devitt2016}; dynamical decoupling protocols that mitigate decoherence \cite{pokharel2018}; simulations of a one-dimensional Ising spin chain \cite{CerveraLierta2018exactisingmodel}; detector tomography \cite{chen2019}; and protocols for entanglement purification and swapping \cite{behera_demonstration_2019}, to name a few.

Experiments in present-day quantum computers can be thought of as analogous to experiments in more traditional contexts such as classical optics. Consider a generic interferometric setup, where laser light's polarization and propagation path are the two qubits of interest. To prepare a target state, wave plates, polarizers, and other instruments transform the beam's polarization. Their action is far from ideal, for they produce unwanted reflections, reduce the beam's power, and increase the state's uncertainty degree, which grows with the number of instruments and their precision. Upon encountering a beam splitter, the beam separates into two paths, whose lengths experience irregular changes due to mechanical perturbations. Both beams undergo more transformations, are reflected on mirrors, which introduce additional phase shifts, and are finally recombined in a beam splitter. The extent to which the resulting state resembles the target state depends on the magnitudes of the alluded factors. With regard to IBM's quantum devices, the state of transmon qubits also suffers from environmental effects, albeit on a much shorter time scale, and from imperfect gate execution and measurement readout. At the beginning of a run of a circuit, each qubit is initialized in the ground state $|0\rangle$. One-qubit (two-qubit) gates are then applied, with error rates ranging from 0.02\% to 0.06\% (0.6\% to 1\%) and a time duration of a few tens (hundreds) of nanoseconds. {Measurement results are read in slightly less than one thousand nanoseconds, with a small but non-negligible error probability between 1\% and 14\%.} The decoherence times $T_1$ and $T_2$ of each qubit range from tens to hundreds of {microseconds}. As in the optical case, the final state differs from the target state, but is found on average to resemble it depending, largely, on the total time required to run the circuit. Because this analogy suggests that the two contexts allow for the experimental realization of similar processes with comparable degrees of uncertainty, we carried out our proposal in the latter.

This paper is organized as follows. We introduce our proposal in Sec.~\ref{sec:realization}, and describe its implementation in an optical as well as in a quantum computational setting. Our results, presented in Sec.~\ref{sec:results}, provide experimental evidence that strong measurements are not universally better than their weaker counterparts, and attest to the usefulness of cloud-based quantum computation devices for experimental tests. We conclude in Sec.~\ref{sec:conclusions} by contrasting our results to Denkmayr \textit{et~al.}'s and placing our work in the broader context.

\section{\label{sec:realization}Weak value measurements}

Consider two interacting qubits, $a$ and $b$. Let $\left\{|0_a\rangle,|1_a\rangle\right\}$ be the basis of qubit $a$, the system, and $\left\{|0_b\rangle,|1_b\rangle\right\}$ the basis of qubit $b$, the pointer. Measurements of the system operator $\sigma^a_{\mathbf{n}}$ are described by $U(\epsilon)=\exp\left(-i\epsilon \sigma^a_{\mathbf{n}} \sigma^b_z \right)$, where $\epsilon$ is a positive number that quantifies the strength of the measurement, $\mathbf{n}$ is a real unit vector, and
\begin{equation}
\begin{gathered}
\sigma^a_{\mathbf{n}}=\mathbf{n}\cdot\bm{\sigma}^a, \quad \bm{\sigma}^a=\left(\sigma^a_x,\sigma^a_y,\sigma^a_z\right), \\
\sigma^a_x=|0_a\rangle\langle1_a|+|1_a\rangle\langle0_a|, \\
\sigma^a_y=-i|0_a\rangle\langle1_a|+i|1_a\rangle\langle0_a|, \\
\sigma^a_z=|0_a\rangle\langle0_a|-|1_a\rangle\langle1_a|,
\end{gathered}
\end{equation}
with $\sigma^b_{x,y,z}$ defined analogously. Weak measurements allow us to approximate $U(\epsilon)$ as $\sigma^a_0\sigma^b_0- i\epsilon \sigma^a_{\mathbf{n}} \sigma^b_{z}+ \mathcal{O}(\epsilon^2)$, {where $\sigma^{a,b}_0$ is the identity operator in each qubit space.} It is possible, however, to go beyond the weak regime. As described in \cite{ruelas2023, deZela22}, $U(\epsilon)$ can be expressed in closed form,
\begin{equation}
    \label{eq:closedFormEvolution}
    U(\epsilon)=\cos\epsilon\, \sigma^a_0\sigma^b_0 -i \sin\epsilon\, \sigma^a_{\mathbf{n}} \sigma^b_{z}.
\end{equation}
{With an eye on the specific proposals to be discussed later, we rewrite the evolution operator as
\begin{align}\label{eq:operatorForImplementation}
     &\left(\cos\epsilon\, \sigma^a_0 \!-\! i\sin\epsilon\, \sigma^a_{\mathbf{n}}\right)|0_b\rangle\langle0_b| \!+\! \left(\cos\epsilon\, \sigma^a_0 \!+\! i\sin\epsilon\, \sigma^a_{\mathbf{n}}\right)|1_b\rangle\langle1_b| \nonumber\\
		        &=\exp\left(-i\epsilon\sigma^a_{\mathbf{n}}\right)|0_b\rangle\langle0_b|+\exp\left(i\epsilon\sigma^a_{\mathbf{n}}\right) |1_b\rangle\langle1_b| \nonumber\\
          &\equiv U^\dagger_a(\mathbf{n}, \epsilon)\Pi_{0_b}+U_a(\mathbf{n}, \epsilon)\Pi_{1_b}.
\end{align}}

The prescription for measuring the weak value $\left\langle \sigma^a_{\mathbf{n}} \right\rangle_w$ goes as follows: first, prepare the preselection (system) state $|A_i\rangle$ and couple it to the (pointer) state $|B_i\rangle=|+_b\rangle$, with $|\pm_b\rangle\equiv2^{-1/2}(|0_b\rangle\pm|1_b\rangle)$. Then, apply the evolution operator in Eq.~(\ref{eq:closedFormEvolution}) to obtain
\begin{align}
    U(\epsilon)|A_i\rangle|B_i\rangle&= \left(\cos\epsilon\, \sigma^a_0\sigma^b_0-i\sin\epsilon \, \sigma^a_{\mathbf{n}}\sigma^b_z\right)|A_i\rangle|+_b\rangle \nonumber\\
   &=\cos\epsilon|A_i\rangle|+_b\rangle-i\sin\epsilon\,\sigma^a_{\mathbf{n}}|A_i\rangle|-_b\rangle.
\end{align}
Afterwards, project the evolved state onto $|A_f\rangle|B_f\rangle$, the postselection state coupled to an auxiliary pointer state, and rewrite the resulting expression as
\begin{align}
    &\langle A_f|\langle B_f|U(\epsilon)|A_i\rangle|B_i\rangle \nonumber\\
    &=\cos\epsilon\langle A_f|A_i\rangle\langle B_f|+_b\rangle-i\sin\epsilon\big\langle A_f\big|\sigma^a_{\mathbf{n}}\big|A_i\big\rangle\langle B_f|-_b\rangle \nonumber\\
    &=\langle A_f|A_i\rangle\left(\cos\epsilon\langle B_f|+_b\rangle-i\sin\epsilon\left\langle\sigma^a_{\mathbf{n}}\right\rangle_w\langle B_f|-_b\rangle\right).\label{eq:finalState}
\end{align}
The quantity in Eq.~(\ref{eq:finalState}) is a complex number, as is the weak value of interest, which we can write in terms of its real and imaginary parts:
\begin{equation}
\left\langle \sigma^a_{\mathbf{n}} \right\rangle_w=R+iI.
\end{equation}
The square of the norm of Eq.~(\ref{eq:finalState}) gives a measurable quantity, from which we can extract all required information by selecting $|B_f\rangle$ to be
\begin{subequations}\label{eq:theoreticalObs}
	\begin{align}
		    &|+_b\rangle\!\longrightarrow i_0=|\langle A_f|A_i\rangle|^2 \cos^2\!\epsilon,\label{eq:observableI0} \\
      &|-_b\rangle\!\longrightarrow i_1 =|\langle  A_f|A_i \rangle|^2 \sin^2\!\epsilon\, \left|\left\langle \sigma^a_{\mathbf{n}} \right\rangle_w\right|^2,\label{eq:observableI1} \\
			&|0_b\rangle\longrightarrow i_2=\frac{1}{2}|\langle A_f|A_i\rangle|^2\left[\left(\cos\epsilon+I\sin\epsilon\right)^2+R^2\sin^2\!\epsilon \right], \label{eq:observableI2}\\
			    &|1_b\rangle\longrightarrow i_3=\frac{1}{2}|\langle A_f|A_i\rangle|^2\left[\left(\cos\epsilon-I\sin\epsilon\right)^2+R^2\sin^2\!\epsilon \right], \label{eq:observableI3}\\
			        &|r_b\rangle\longrightarrow i_4=\frac{1}{2}|\langle A_f|A_i\rangle|^2\left[\left(\cos\epsilon+R\sin\epsilon\right)^2+I^2\sin^2\!\epsilon \right], \label{eq:observableI4}\\
				    &|l_b\rangle\longrightarrow i_5=\frac{1}{2}|\langle A_f|A_i\rangle|^2\left[\left(\cos\epsilon-R\sin\epsilon\right)^2+I^2\sin^2\!\epsilon \right]\label{eq:observableI5},
	\end{align}
\end{subequations}
with
\begin{equation*}
|r_b\rangle\equiv2^{-1/2} (|0_b\rangle +i|1_b \rangle) \,\,\,\text{and}\,\,\, |l_b\rangle \equiv2^{-1/2} (|0_b\rangle-i|1_b\rangle).
\end{equation*}
We call $i_{0,\ldots,5}$ intensities, regardless of whether they represent laser power, photon coincidence counts, or other physical quantities. Subtracting Eqs.~(\ref{eq:observableI3}) and (\ref{eq:observableI5}) from Eqs.~(\ref{eq:observableI2}) and (\ref{eq:observableI4}), respectively, yields
\begin{subequations}\label{eq:subtractionBoth}
\begin{align}
			&i_2-i_3=I|\langle A_f|A_i\rangle|^2\sin2\epsilon, \label{eq:subtractioni2i3}\\
			&i_4-i_5=R|\langle A_f|A_i\rangle|^2\sin2\epsilon\label{eq:subtractioni4i5}.
\end{align}
\end{subequations}
When the overlap $\langle A_f|A_i\rangle$ is non-zero and $\sin2\epsilon\neq0$, {it is possible to solve Eqs.~(\ref{eq:observableI0}), (\ref{eq:observableI1}), (\ref{eq:subtractioni2i3}), and (\ref{eq:subtractioni4i5}) for the squared magnitude, real part, and imaginary part of $\left\langle\sigma^a_{\mathbf{n}}\right\rangle_w$:
\begin{equation}\label{eq:magnitudeRealImaginary}
	    \left|\left\langle \sigma^a_{\mathbf{n}} \right\rangle_w\right|^2 = \frac{i_1}{i_0} \cot^2\!\epsilon,\,      R=\frac{i_4-i_5}{2\, i_0} \cot\epsilon,\,    I=\frac{i_2-i_3}{2\, i_0}\cot\epsilon.
\end{equation}
The above equations are periodic in $\epsilon$ with period $\pi$. We emphasize that, while these results hold for both weak and strong couplings, the values $\epsilon=0,\pi/2,$ and $\pi$ are exceptional, as will be seen in Sec.~\ref{sec:results}.}

{\subsection{\label{sec:opticalProposal}All-optical realization}}

\begin{figure}[ht!]
\centering\includegraphics[width=\columnwidth]{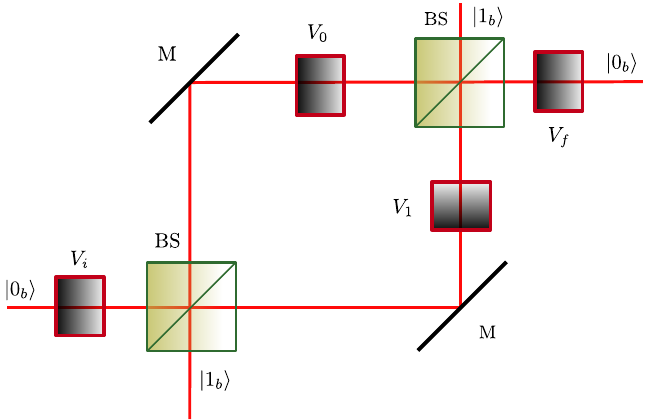}
\caption{\label{fig:interferometerProposal}Mach-Zehnder type interferometer with mirrors (M), beam splitters (BS), and unitary operators $V_{0, 1, i, f}$ for implementing the evolution operator $U(\epsilon)$. The horizontal (vertical) path corresponds to the state $|0_b\rangle$ $\left(|1_b\rangle\right)$.}
\end{figure}

To implement the weak value measurement proposal in an optical setting, with either classical or quantum light, consider the Mach-Zehnder type interferometer shown in Fig~\ref{fig:interferometerProposal}. Let qubit $a$ represent the polarization of light---so that $|0_a\rangle$ and $|1_a\rangle$ represent horizontally and vertically polarized light, respectively---and qubit $b$ its propagation direction, with $|0_b\rangle$ representing the horizontal and $|1_b\rangle$ the vertical direction. {Incoming light} on the horizontal path undergoes a transformation given by the operator $V_i$, whereas incoming light on the vertical path experiences no transformation. Both beams then enter the interferometer through a 50:50 beam splitter (BS), are reflected on mirrors (M), and are acted upon by $V_0$ in the horizontal path and by $V_1$ in the vertical path. Subsequently, they are recombined in another BS, and the beam in the horizontal path undergoes one last transformation, $V_f$. The evolution operator given by Eq.~(\ref{eq:operatorForImplementation}) can be realized in such an interferometer with
\begin{equation*} V_{0}=V_1=U_a(\mathbf{n},\epsilon) \,\,\,\text{and}\,\,\, V_i=V_f=U_a^{\dagger}(\mathbf{n},\epsilon),
\end{equation*}
where $U_a(\mathbf{n},\epsilon)$ is a sequence of quarter-wave ($Q$) and half-wave ($H$) plates,
\begin{align}\label{eq:transformationPlates}
  U_a(\mathbf{n}, \epsilon)=& Q\!\left(\frac{\pi+\varphi}{2}\right)Q\!\left(\frac{\vartheta+\varphi}{2}\right)H\!\left(\frac{-\pi+\vartheta+\varphi}{2}+\frac{\epsilon}{2}\right)\nonumber\\
  &\times Q\!\left(\frac{\vartheta+\varphi}{2}\right)Q\!\left(\frac{\varphi}{2}\right),
\end{align}
the unit vector is $\mathbf{n}=(\sin\vartheta\cos\varphi, \sin\vartheta\sin\varphi, \cos\vartheta)$, and the argument of each plate represents the angle made by its major axis with respect to the vertical direction. The adjoint operator, $U_a^\dagger(\mathbf{n}, \epsilon)$, is given by the same plates in reverse order and with an additional $+\pi/2$ in each argument. The initial and final system-pointer states, $|A_i \rangle |B_i\rangle$ and $|A_f\rangle|B_f\rangle$, respectively, can be implemented by using similar interferometric arrangements at both the input and output ports of the interferometer shown in Fig~\ref{fig:interferometerProposal}. (See \cite{ruelas2023}, \cite{englert2001}, and \cite{[][{ (to be published).}]deZela2022proceedings} for more details on this setup.) Finally, intensities and photon counts can be measured at the output ports, whereupon weak values can be ascertained through Eqs.~(\ref{eq:magnitudeRealImaginary}).

\subsection{\label{sec:computationalProposal}Quantum computational realization}

\begin{figure}[ht!]
\centering\includegraphics[width=\columnwidth]{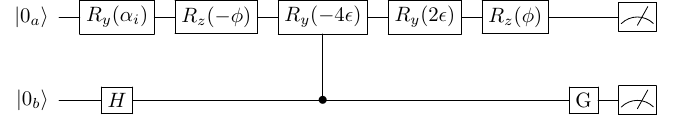}
    \caption{\label{fig:circuit1}Quantum circuit for measuring $\left\langle \sigma^a_{\mathbf{n}} \right\rangle_w$ with interaction strength given by $\epsilon$. It prepares the initial state $|A_i\rangle|B_i\rangle$ and acts $U(\epsilon)$ on it. Then, $i_{0, \ldots, 5}$ are measured by choosing the appropriate gates G, followed by projections onto $\langle0_a|\langle0_b|$ and $\langle0_a|\langle1_b|$ (see text).}   
\end{figure}
To implement the proposal in a quantum computing system, consider, for instance, a Pauli spin operator given by $\mathbf{n}=(-\sin\phi,\cos\phi, 0)$. The operators $\exp\left(\pm i\epsilon\sigma^a_{\mathbf{n}}\right)$ can be decomposed into a product of rotations:
\begin{align}\label{eq:exponentialRotations}
	    \exp\left(\pm i\epsilon\sigma^a_{\mathbf{n}}\right)&=\cos\epsilon\pm i\sin\epsilon\left(-\sin\phi\,\sigma_x^a+\cos\phi\,\sigma_y^a\right) \nonumber \\
	        &=\exp(-i\phi\sigma_z^a/2)\exp(\pm i\epsilon\sigma_y^a)\exp(i\phi\sigma_z^a/2) \nonumber\\
		    &\equiv R_z(\phi)R_y(\mp 2\epsilon)R_z(-\phi).
\end{align}
With $\exp\left(\pm i\epsilon \sigma^a_{\mathbf{n}} \right)$ given by Eq.~(\ref{eq:exponentialRotations}), the evolution operator $U(\epsilon)$ of Eq.~(\ref{eq:operatorForImplementation}) can be realized in the quantum circuit shown in Fig.~\ref{fig:circuit1}. The system state is prepared as the desired preselection state by applying a rotation $R_y$ to the state $|0_a\rangle$: $|A_i\rangle =R_y(\alpha_i)|0_a\rangle =\cos(\alpha_i/2) |0_a\rangle+ \sin(\alpha_i/2) |1_a\rangle$. The pointer state, in turn, is set to the auxiliary state $|B_i\rangle= |+_b\rangle=H|0_b\rangle$, where $H$ is the Hadamard gate,
\begin{align}
	\label{eq:hadamardDef}
	H&=\frac{1}{\sqrt{2}}\left(|0_b\rangle\langle0_b|+|0_b\rangle\langle1_b|+|1_b\rangle\langle0_b|-|1_b\rangle\langle1_b|\right) \nonumber\\
 &=|+_b\rangle\langle0_b|+|-_b\rangle\langle1_b|.
\end{align}
To optimize the number of gates employed in the circuit, we substitute the exponentials of Eq.~(\ref{eq:exponentialRotations}) into Eq.~(\ref{eq:operatorForImplementation}) and regroup them as follows:
\begin{align}\label{eq:operatorForImplementation3}
	U(\epsilon)&=R_z(\phi)R_y( 2\epsilon)R_z(-\phi)\Pi_{0_b} \nonumber\\
 &\quad +R_z(\phi)R_y(-2\epsilon)R_z(-\phi)\Pi_{1_b}\nonumber\\
	&= R_z(\phi)R_y( 2\epsilon)\mathrm{C}\!R_y(-4\epsilon)R_z(-\phi),
\end{align}
{where we have defined the controlled $R_y$ gate,
\begin{equation*}
\mathrm{C}\!R_y(\vartheta)\equiv \Pi_{0_b}+ R_y(\vartheta) \Pi_{1_b},
\end{equation*}
which applies a rotation on qubit $a$ depending on the state of $b$.} The evolution operator first subjects qubit $a$ to the rotation $R_z(-\phi)$. Then, if the state of qubit $b$ is $|0_b\rangle$, $a$ is acted on by $R_y(2\epsilon)$; otherwise, $R_y(-4\epsilon)$ followed by $R_y(2\epsilon)$ are applied on qubit $a$. Lastly, $R_z(\phi)$ acts on $a$. To project the evolved state onto $\langle A_f|\langle B_f|$, we choose {the postselection state $|A_f\rangle=|0_a\rangle$, let different gates G act on qubit $b$ [in order to select the different $|B_f\rangle$ of Eqs.~(\ref{eq:theoreticalObs})], and measure both qubits, as indicated in Fig.~\ref{fig:circuit1}.} The observables $i_{0, 1}$ result from applying a Hadamard gate to qubit $b$, i.e. $\mathrm{G}=H$, and projecting onto $\langle0_a|\langle0_b|$ and $\langle0_a|\langle1_b|$, respectively. To measure $i_{2, 3}$, $\mathrm{G}$ is set equal to the identity and is followed by the same respective projections. For $i_{4, 5}$ we set $\mathrm{G} =H\,S^{\dagger}$, with
\begin{equation}
	\label{eq:sdagDef}
	S^{\dagger} =|0_b\rangle\langle0_b|+ e^{-i\pi/2} |1_b\rangle\langle1_b| =|0_b\rangle\langle0_b|-i|1_b\rangle\langle1_b|.
\end{equation}
Thus, $S^{\dagger}$ introduces a relative phase shift of $-\pi/2$ between $|0_b\rangle$ and $|1_b\rangle$. The intensities, again, are obtained by projecting onto $\langle0_a|\langle0_b|$ and $\langle0_a|\langle1_b|$.

{Moreover,} knowledge of $\left\langle \sigma^a_{\mathbf{n}} \right\rangle_w$ allows us to determine the preselection state \cite{lundeen2011, lundeen2012, salvail2013, maccone2014}. Any Pauli spin operator has two eigenvectors,
\begin{equation}
\label{eq:eigensystemPauli}
\sigma^a_{\mathbf{n}} |\mathbf{n}_\pm\rangle=\pm |\mathbf{n}_\pm\rangle,
\end{equation}
whose eigenprojectors can be expressed as
\begin{equation}\label{eq:eigenprojectorsSigma}
    \Pi_\pm^a=|\mathbf{n}_\pm \rangle \langle \mathbf{n}_\pm|=\frac{1}{2} \left( \sigma^a_0\pm\sigma^a_{\mathbf{n}} \right).
\end{equation}
For our choice of $\mathbf{n}$, preselection, and postselection states, the theoretical real and imaginary parts of the weak value are given by
\begin{equation}\label{eq:weakValueTheo}
\left\langle\sigma^a_{\mathbf{n}}\right\rangle_w =R_{t}+iI_{t} \equiv-\tan\! \left(\alpha_i/2\right) (\sin\phi+i\cos\phi),
\end{equation}
the eigenvectors of $\sigma^a_\mathbf{n}$ are found to be
\begin{equation}\label{eq:eigenvectorsSigma}
    |\mathbf{n}_\pm\rangle=\frac{1}{\sqrt{2}}\left(\pm|0_a\rangle+ie^{i\phi}|1_a\rangle\right),
\end{equation}
and the weak values of the eigenprojectors in Eq.~(\ref{eq:eigenprojectorsSigma}) take the form
\begin{equation}
    \label{eq:weakValuesEigenprojectors}
    \left\langle \Pi_\pm^a \right\rangle_w =\frac{1}{2} \left(1 \pm \left\langle \sigma^a_{\mathbf{n}} \right\rangle_w \right) = \frac{1}{2}\left( 1\mp ie^{-i\phi}\tan(\alpha_i/2)\right).
\end{equation}
With the goal of comparing our results to Denkmayr \textit{et~al.}'s \cite{denkmayr2017}, we now describe how to characterize the preselection state's parameters. The components of $|A_i\rangle$ in the basis of $\sigma^a_{\mathbf{n}}$ are
\begin{equation}
    \langle\mathbf{n}_\pm|A_i\rangle= \pm\frac{1}{\sqrt{2}} \left(\cos(\alpha_i/2)\mp ie^{-i\phi}\sin(\alpha_i/2)\right),
\end{equation}
so the preselection state can be written as
\begin{equation}\label{eq:measuredPreselection}
    |A_i\rangle =\frac{\big\langle \Pi_+^a \big\rangle_w |\mathbf{n}_+\rangle -\big\langle \Pi_-^a \big\rangle_w |\mathbf{n}_-\rangle}{\sqrt{\big|\big\langle \Pi_+^a \big\rangle_w \big|^2+ \big|\big\langle \Pi_-^a \big\rangle_w \big|^2}}.
\end{equation}
The relative minus sign in Eq.~(\ref{eq:measuredPreselection}) is a consequence of having
\begin{equation}
\label{eq:postselectionFinal}
|A_f\rangle=|0_a\rangle=\frac{1}{\sqrt{2}}\left(|\mathbf{n}_+\rangle-|\mathbf{n}_-\rangle\right)
\end{equation}
instead of
\begin{equation*} 
|1_a\rangle=\frac{1}{\sqrt{2}} e^{-i(\phi+\pi/2)}  \left(|\mathbf{n}_+\rangle +|\mathbf{n}_-\rangle\right)
\end{equation*}
as postselection state \footnote{The state $|1_a\rangle$ is, up to a global phase, the ``diagonal'' state of $\sigma_\mathbf{n}^a$, just as the postselection state in \cite{denkmayr2017}, $|P_x;+\rangle= 2^{-1/2}\left( |P_z;+\rangle +|P_z;-\rangle\right)$, was the diagonal state of $\sigma_z$ (cf. Eqs.~(5) and (8) of \cite{denkmayr2017}). ``Diagonal'' and ``antidiagonal'' states (e.g., $|1_a\rangle$ and $|0_a\rangle$) and the basis states ($|\mathbf{n}_+\rangle$ and $|\mathbf{n}_-\rangle$) are said to form mutually unbiased bases, of which the ``diagonal'' states are special members (see, for instance, \cite{lundeen2011, salvail2013}).}. Then, we define the normalization factor
\begin{equation}\label{eq:normalizationFactor}
    \nu\! \equiv\! \left(\big|\big\langle \Pi_+^a \big\rangle_w \big|^2 \! + \!\big|\big\langle \Pi_-^a \big\rangle_w \big|^2 \right)^{-1/2}\!=\! \left(\frac{1}{2} \! + \! \frac{1}{2}\big|\big\langle \sigma^a_{\mathbf{n}} \big\rangle_w\big|^2 \right)^{-1/2}\!\!.
\end{equation}
For our choice of states and unit vector, Eq.~(\ref{eq:normalizationFactor}) yields the theoretical normalization factor $\nu_t=\sqrt{2}\cos(\alpha_i/2)$. Therefore, the characterization of Eq.~(\ref{eq:measuredPreselection}) [or, equivalently, of Eqs.~(\ref{eq:weakValuesEigenprojectors}) and (\ref{eq:normalizationFactor})] is achieved by compounding Eqs.~(\ref{eq:theoreticalObs}) as
\begin{subequations}\label{eq:finalPrescription}
\begin{align}
    \nu&= \left(\frac{1}{2}+\frac{1}{2}\frac{i_1}{i_0} \cot^2 \!\epsilon\right)^{-1/2}\!, \\
    \alpha&= 2\arctan \sqrt{\frac{i_1}{i_0} \cot^2 \! \epsilon}, \\
    \phi&= \arctan\! \left(\frac{i_4-i_5}{i_2-i_3}\right).
\end{align}
\end{subequations}
Equations~(\ref{eq:magnitudeRealImaginary}) and (\ref{eq:finalPrescription}) express all the parameters of interest in terms of the measurable quantities $i_{0,\ldots,5}$. We employed them to obtain the results presented in the following section.

\section{\label{sec:results}Results}

\begin{figure*}[ht!]\centering
    \includegraphics[width=0.9\textwidth]{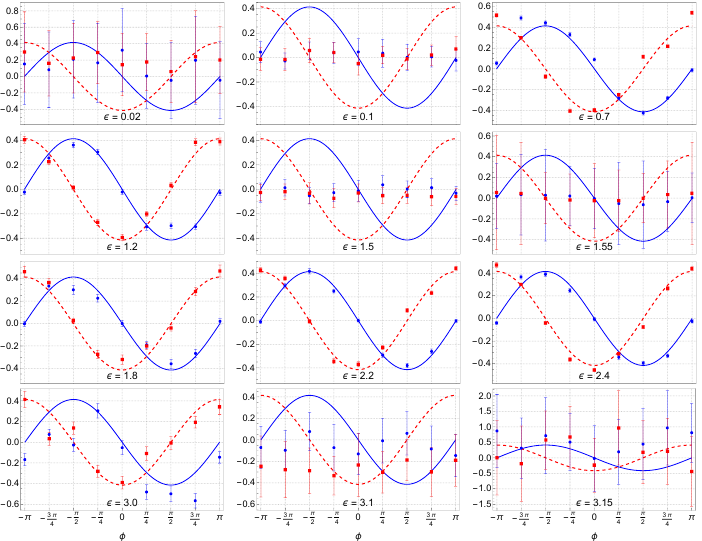}
    \caption{Weak values $\left\langle\sigma^a_{\mathbf{n}}\right\rangle_w$ obtained in ibm\_oslo as functions of the rotation angle $\phi$, with $\alpha_i=\pi/4$, for twelve measurement strengths $\epsilon$. The solid blue lines (circles) stand for the theoretical (experimental) real part of the weak value, $R_t$ ($R_m$). The dashed red lines (squares) represent the theoretical (experimental) imaginary part of the same weak value, $I_t$ ($I_m$). Error bars represent $\sigma_{R}$ or $\sigma_{I}$ [see Eqs.~(\ref{eq:errorPropagation})].}
    \label{fig:weakResults}
\end{figure*}
\begin{figure}[ht!]
    \centering
    \includegraphics[width=\columnwidth]{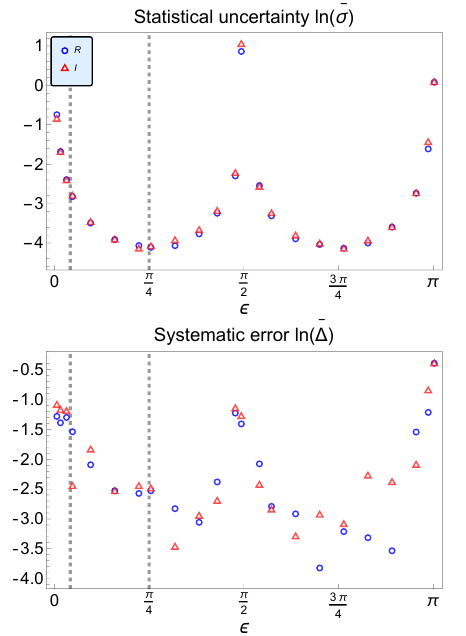}
    \caption{{Logarithmic plots of $\bar{\sigma}$, a measure of statistical uncertainty for assessing precision, and $\bar{\Delta}$, a metric for systematic errors that reflects accuracy,} as defined in Eq.~(\ref{eq:resultprec}), of the real (blue circles) and imaginary (red triangles) parts of $\big\langle \sigma^a_{\mathbf{n}} \big\rangle_w$, as functions of the measurement strength $\epsilon$. The vertical dotted lines represent $\epsilon=\pi/24$ and $\pi/4$, the values analyzed by Denkmayr \textit{et al.} \cite{denkmayr2017}.}
    \label{fig:precAndAcc}
\end{figure}

\begin{figure*}[ht!]\centering
    \includegraphics[width=0.9\textwidth]{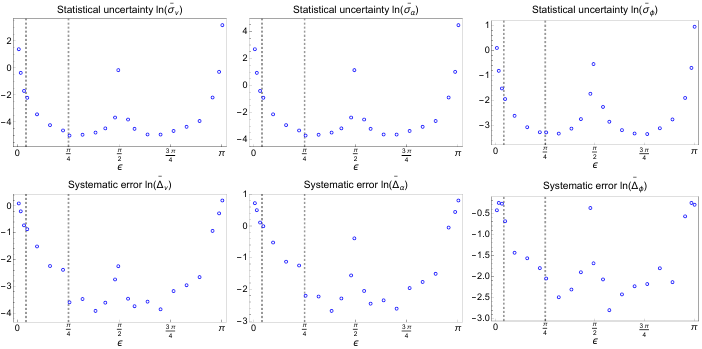} 
    \caption{{Statistical uncertainty} (top) and systematic error (bottom), as quantified by $\ln\!\left(\bar{\sigma}\right)$ and $\ln\!\left(\bar{\Delta}\right)$, respectively, of the normalization factor $\nu$ (left), amplitude angle $\alpha$ (center), and phase $\phi$ (right), as functions of the measurement strength $\epsilon$. The vertical dotted lines represent $\epsilon=\pi/24$ and $\pi/4$, the values analyzed by Denkmayr \textit{et al.} \cite{denkmayr2017}.}
    \label{fig:directMeasurementResults}
\end{figure*}

Our work was conducted in ``ibm\_oslo'', an open-access quantum system with seven qubits, labeled 0 to 6, from which we chose qubits 0 and 1 to carry out our experiments. During data recollection, qubit 0 (1) had an average $T_1$ of 142 $\mu$s (135 $\mu$s) and an average $T_2$ of 101 $\mu$s (26 $\mu$s) \footnote{Preliminary tests revealed that ibm\_oslo, as well as other open-access quantum systems, produced data similar to that reported in Sec.~\ref{sec:results} whenever their relaxation times differed by a few tens of microseconds from our average values. These parameters sometimes dropped significantly, however. As a consequence, the multiple error sources inherent to the hardware rendered all measurement outcomes meaningless.}. As explained in Sec.~\ref{sec:realization}, three different circuits are required to compute $\left\langle\sigma^a_{\mathbf{n}}\right\rangle_w$; their outcomes are a measurement of either $i_0$ and $i_1$, $i_2$ and $i_3$, or $i_4$ and $i_5$. For each value of $\epsilon$ and $\phi$ sampled, we ran each circuit 2000 times, repeated this procedure 20 times, averaged the intensities, and computed the parameters of interest in accordance with Eqs.~(\ref{eq:magnitudeRealImaginary}) and (\ref{eq:finalPrescription}).

{To distinguish between theoretical and experimental parameters, let us denote the theoretical intensities as ${i_{0t},\ldots, i_{5t}}$, which are given by Eqs.~(\ref{eq:theoreticalObs}) with $|\langle A_f|A_i\rangle|^2\to \cos^2\! \left(\alpha_i/2\right)$, $I\to I_{t}$, and $R\to R_{t}$. The experimental mean values of these parameters will be denoted as $R_{m}$ and $I_{m}$.} Each mean intensity has an associated standard deviation $\sigma_{i_0}, \ldots, \sigma_{i_5}$. A measure of uncertainty for the experimental weak values is given by the propagation formulae
\begin{subequations}\label{eq:errorPropagation}
\begin{align}
	\sigma_{R} &=\sqrt{\left( {\frac{\partial R}{\partial i_0}}\right)^2 \sigma_{i_0}^2+ \left(\frac{\partial R}{\partial i_4}\right)^2 \sigma_{i_4}^2+ \left( \frac{\partial R}{\partial i_5}\right)^2 \sigma_{i_5}^2} \nonumber\\
 &= \frac{1}{\cos^2\! \left(\alpha_i/2\right)} \sqrt{\frac{\tan^2\!\left(\alpha_i/2\right) \sin^2\!\phi}{\cos^4\!{\epsilon}}\sigma_{i_0}^2+\frac{\sigma_{i_4}^2 + \sigma_{i_5}^2}{\sin^2\!{2\epsilon}}}, \label{eq:sigmaReal}\\ 
    \sigma_{I} &=\sqrt{\left(\frac{\partial I}{\partial i_0}\right)^2\sigma_{i_0}^2+ \left(\frac{\partial I}{\partial i_2}\right)^2\sigma_{i_2}^2+ \left(\frac{\partial I}{\partial i_3}\right)^2\sigma_{i_3}^2} \nonumber\\
 &= \frac{1}{\cos^2\! \left(\alpha_i/2\right)} \sqrt{\frac{\tan^2\!\left(\alpha_i/2\right) \cos^2\!\phi}{\cos^4\!{\epsilon}}\sigma_{i_0}^2+\frac{\sigma_{i_2}^2 + \sigma_{i_3}^2}{\sin^2\!{2\epsilon}}}, \label{eq:sigmaIm}
\end{align}
\end{subequations}
where the derivatives are evaluated at the theoretical intensities. Similar relations hold for the uncertainties of the preselection state parameters, $\sigma_\nu, \sigma_\alpha, \sigma_\phi$.

For our experiments, we swept the measurement strength $\epsilon$ across a full period. At each value of $\epsilon$, we took $N=9$ samples of $\phi$ from $-\pi$ to $\pi$, and set $\alpha_i=\pi/4$. Following Denkmayr \textit{et al.} \cite{denkmayr2017}, we introduce figures of merit for our results at fixed $\epsilon$. Let $\mu$ refer to $R$, $I$, $\nu$, $\alpha$, or $\phi$, so that $\mu_{t, j}$ is given by Eq.~(\ref{eq:weakValueTheo}) evaluated at $\phi_j$, or by
\begin{subequations}
	\label{eq:sampledParameters}
	\begin{align}
		\nu_{t, j}&=\sqrt{2}\cos(\alpha_i/2), \label{eq:sampledNormalization}\\
		\alpha_{t, j}&=\pi/4, \label{eq:sampledAlpha}\\
		\phi_{t, j}&\in\{-\pi, -3\pi/4, \ldots, 3\pi/4, \pi\}. \label{eq:sampledPhi}
	\end{align}
\end{subequations}
Similarly, $\sigma_{\mu, j}$ refers to Eqs.~(\ref{eq:errorPropagation}) or $\sigma_\nu, \sigma_\alpha, \sigma_\phi$, and $\mu_{m, j}$ is the mean value of $\mu$ obtained for sample $j$. Using this notation, the achieved degrees of precision and accuracy, respectively, can be assessed by means of the following parameters:
\begin{equation}
	\label{eq:resultprec}
	    \bar{\sigma}_\mu\equiv\sqrt{\frac{1}{N}\sum_j^N\sigma_{\mu, j}^2},\quad\bar{\Delta}_\mu\equiv\sqrt{\frac{1}{N}\sum_j^N\left(\mu_{t, j}-\mu_{m, j}\right)^2}.
\end{equation}
For a given strength $\epsilon$, $\bar{\sigma}_\mu$ is the root mean square (RMS) value of the sample uncertainties. It serves as a proxy for precision, in that it quantifies the statistical uncertainty of the sample. Likewise, $\bar{\Delta}_\mu$ is the RMS value of the difference between theoretical and experimental parameters in a sample. So defined, it appraises the accuracy of our results by representing systematic errors in the data.

Figure~\ref{fig:weakResults} shows the real and imaginary parts of $\left\langle\sigma^a_{\mathbf{n}}\right\rangle_w$ as functions of the rotation angle $\phi$, for twelve different measurement strengths. For $\epsilon$ near zero, $R_{m}$ and $I_{m}$ result from subtracting two very close numbers [see Eqs.~(\ref{eq:observableI2})-(\ref{eq:observableI5})] and dividing by $\sin2\epsilon$, which also tends to zero. The experimental results behave accordingly: both parameters fluctuate about 0, with greater dispersion and error for smaller $\epsilon$. As measurements become stronger, the results are in better accordance with theoretical predictions, and fluctuations subside. When $\epsilon$ approaches $\pi/2$, this trend reverses, with the weak values behaving similarly to how they did close to $\epsilon=0$. For larger values of $\epsilon$, the results improve until $\epsilon$ nears $\pi$, at which point the previous erratic behavior re-emerges. Hence, critical values in the regime of strong coupling exist, for which precision and accuracy behave the same as in the weak coupling regime.

Figures~\ref{fig:precAndAcc} and \ref{fig:directMeasurementResults} summarize the information contained in our results that pertains to statistical uncertainty and systematic errors. In Fig.~\ref{fig:precAndAcc}, $\bar{\sigma}_{R, I}$ take relatively high values for very weak measurements, but decrease as $\epsilon$ grows up to $\pi/4$, where they begin to increase. Since $\bar{\sigma}_{R, I}$ are the RMS values of the uncertainties in Eqs.~(\ref{eq:errorPropagation}), which diverge for $\epsilon=0,\pi/2$, and $\pi$, we can expect $\bar{\sigma}_{R, I}$ to reach relative maxima at these strengths. This prediction is confirmed by Fig.~\ref{fig:precAndAcc}. The height of each maximum of $\bar{\sigma}_{R, I}$ depends on both the standard deviations and their respective coefficients in Eqs.~(\ref{eq:errorPropagation}). In the rest of the domain of $\epsilon$, $\bar{\sigma}$ first drops and then rises until the last divergence is reached. On the other hand, $\bar{\Delta}$ reflects how much the measured values differ from the theoretical predictions. As we saw in Fig.~\ref{fig:weakResults} and argued above, the resemblance is worse for $\epsilon$ in the vicinity of $0, \pi/2$, and $\pi$ because Eqs.~(\ref{eq:magnitudeRealImaginary}) diverge at these points. The results in Fig.~\ref{fig:precAndAcc} illustrate this trend, with $\bar{\Delta}$ first decreasing and then rising in between divergences, albeit less smoothly than $\bar{\sigma}$. Figure~\ref{fig:directMeasurementResults} displays the same tendencies for the precision and accuracy of the preselection state parameters $\nu$, $\alpha$, and $\phi$ \footnote{In samples where $\tan\phi_j=0$ (for example, $\phi_j=-\pi$) the experimental values of $i_4$ and $i_5$ are two close numbers. Due to random fluctuations, intensities such that $i_4-i_5>0$ or $i_4-i_5<0$ are both found. As a consequence, the inverse tangent function introduces a spurious $\pm\pi$ factor whenever the difference has the wrong sign (positive, in our example). To avoid this issue, we computed the angle $\phi$ as $\arctan\left|\dfrac{i_4-i_5}{i_2-i_3}\right|$, thereby constraining the results to the first quadrant and the theoretical phases $\phi_{t,j}$ to $\left\{0,\pi/4,\pi/2\right\}$.}.

\section{\label{sec:conclusions}Summary and conclusions}

In this work, we have extended Denkmayr \textit{et~al.}'s scheme for analysing the behavior of the weak value $\left\langle\sigma_z  \right\rangle_w$ with measurements of arbitrary strength \cite{denkmayr2017, denkmayr2018}. We tested this extension with the quantum computational tools hosted by IBM. Our results show that the model works as expected: $\left\langle\sigma^a_{\mathbf{n}}\right\rangle_w$ and {the preselection state $|A_i\rangle$} can be computed for all allowed strengths, but the statistical uncertainty and systematic errors present in each reconstruction do not decrease monotonically with increasing strength.

In the paper that inspired the present work, Denkmayr \textit{et~al.} \cite{denkmayr2017} compared precision and accuracy for two measurement strengths, given by $\alpha=\pi/12$ and $\pi/2$ in their notation, which correspond to $\epsilon=\pi/24$ and $\pi/4$ in ours. They concluded that strong measurements perform better than their weaker counterparts. From both Fig.~\ref{fig:precAndAcc} and Fig.~\ref{fig:directMeasurementResults} we can conclude the same for those cases---the difference in the quality of the results is compelling. If statistical uncertainties and systematic errors were monotonically decreasing functions of $\epsilon$, we could further assert that stronger measurements are universally better than weaker ones. It must be stressed, however, that this is not the case. (In the experimental setting of \cite{denkmayr2017}, $\alpha/2=\pi/4$ was the actual maximum measurement strength possible. This may have constrained the analysis.) Figures~\ref{fig:precAndAcc} and \ref{fig:directMeasurementResults} also suggest that it could be possible to minimize statistical uncertainties and systematic errors (equivalently, maximize precision and accuracy) by choosing $\epsilon$ appropriately.

Denkmayr \textit{et al.} remarked that their model ``can be used for any coupling between two two-level quantum systems'' \cite{denkmayr2017, denkmayr2018}. As we outlined in Sec.~\ref{sec:opticalProposal}, our measurement scheme can be implemented in an all-optical interferometric setup that admits classical light beams as well as single photons. {Furthermore, their results \cite{denkmayr2017, denkmayr2018} and those of other authors \cite{calderaro2018, xu2021, vallone2016} seemed to suggest that, in general, stronger measurements are superior to weaker ones in terms of precision and accuracy. Previous theoretical analyses \cite{das2014estimation, gross2015, zhu2016direct} had shown that this need not always be the case. To the best of our knowledge, here we have presented the first experimental observation of such a behavior.}

A recent survey of several IBM quantum systems has shed light on how their performances compare to each other and how their individual qubits compare to one another \cite{patel2020}. In particular, it revealed the degree to which their relaxation times, gate error rates, and readout error rates vary across the sampled machines and over time. As argued throughout this article, we can expect our model to yield similar results when run on different devices---as long as the relaxation times do not dwindle considerably. How the present scheme fares in different experimental contexts and in contrast to, e.g., standard tomography is an open question whose undertaking we would welcome.

\begin{acknowledgments}
The authors acknowledge the use of IBM Quantum services for this work. The views expressed are ours, and do not reflect the official policy or position of IBM or the IBM Quantum team. D. R. A. R. P. acknowledges funding by FONDECYT through Grant 236-2015.
\end{acknowledgments}

\bibliography{biblio}

\end{document}